\documentclass[cits]{PoS}
\newcommand{\psib}{{\overline{\psi}}}
\usepackage{amssymb}
\usepackage{fontenc}
\usepackage{times}
\usepackage{mathptmx}
\usepackage{graphicx}

\title{Fermion mass generation without a condensate}
\ShortTitle{Fermion mass generation without a condensate}

\author{\speaker{Venkitesh Ayyar} \thanks{Work done in collaboration with Shailesh Chandrasekharan. The material presented here is based upon work supported by the U.S. Department of Energy, Office of Science, Nuclear Physics program under Award Number DE-FG02-05ER41368. This research was done using resources provided by the Open Science Grid, which is supported by the National Science Foundation and the U.S. Department of Energy's Office of Science.}\\
Duke University, Durham NC, USA.\\
E-mail: \email{vpa@phy.duke.edu}}

\abstract{We study a lattice field theory model containing two flavors of massless staggered fermions with an onsite four-fermion interaction. The symmetry of the model forbids non-zero fermion bilinear order parameters that can generate a fermion mass. At weak couplings, we expect a massless fermion phase. At strong couplings, we can argue for the existence of massive fermions without the formation of any fermion bilinear condensate. Using Monte Carlo calculations in three space-time dimensions, we find evidence for a direct second order phase transition between the two phases.
}

\FullConference{The 32nd International Symposium on Lattice Field Theory,\\
		23-28 June, 2014\\
		Columbia University New York, NY}

\begin{document}

\section{Introduction}
It is well known that relativistic four-fermion field theories in three dimensions can contain strongly interacting second order fixed points \cite{warr,hands} . The main motivation for their study is to understand new mechanisms for fermion mass generation that may be realized in nature. The conventional mechanism of fermion mass generation is via spontaneous symmetry breaking of chiral symmetries which is signalled through a non-zero fermion bilinear condensate. In this work, we explore a more exotic mechanism of fermion mass generation without such condensates.

We study a simple lattice four-fermion model in three Euclidean dimensions. This model is a limiting case of a Yukawa model. In the weak-coupling regime, since the coupling is irrelevant, we have a massless phase. In the strong coupling regime, we can argue for the existence of a massive phase. Interestingly, all fermion bilinear condensates are found to vanish in both phases. This exotic massive phase with zero condensates has been studied before \cite{anna}. However, a first principles Monte Carlo calculation to study the transition between these phases in 3D had not been done before. In this work, we perform such a calculation and show that these two phases are separated by a single second-order phase transition. A detailed description of this work has been published in \cite{our}.

\section{Model}
We study a simple four-fermion model containing two flavors of staggered fermions whose Euclidean action is given by
\begin{equation}
S = \sum_{i=1,2}\ \sum_{x,y} {\psib}_{x,i} \ M_{x,y} \ \psi_{y,i}\ - U \ \sum_{x} \Big\{ \psib_{x,1}\psi_{x,1}\psib_{x,2}\psi_{x,2}\Big\} 
\label{act}
\end{equation}
where $\psib_{x,i}, \psi_{x,i}, i = 1,2$ are four independent Grassmann valued fields. The matrix $M$ is the well known staggered fermion matrix given by
\begin{equation}
M_{x,y} \ =\  \sum_{\hat{\alpha}} \frac{\eta_{x,{\hat{\alpha}}}}{2}\ [\delta_{x,y+\hat{\alpha}} - \delta_{x,y-\hat{\alpha}}]
\label{staggered}
\end{equation}
where $x \equiv (x_1,x_2,x_3)$ denotes a lattice site on a $3$ dimensional cubic lattice and $\hat{\alpha} = \hat{1},\hat{2},\hat{3}$  represent unit lattice vectors in the three directions. The staggered fermion phase factors are defined as usual: $\eta_{x,\hat{1}}=1,, \eta_{x,\hat{2}}=(-1)^{x_1}$, and $\eta_{x,\hat{3}}= (-1)^{x_1+x_2}$. We will study cubical lattices of equal size $L$ in each direction with anti-periodic boundary conditions.

It  can be shown that, in addition to the usual space-time lattice transformations ( translation, axis reversal, rotation \cite{sym1,sym2} ) , the action is symmetric under internal $SU(4)$ transformations given by :

\begin{equation}
\left(
\begin{array}{c}
\psi_{x_e,1} \cr
\psib_{x_e,1} \cr
\psi_{x_e,2} \cr
\psib_{x_e,2}
\end{array}
\right)
\ \rightarrow \ 
 V \ \left(
\begin{array}{c}
\psi_{x_e,1} \cr
\psib_{x_e,1} \cr
\psi_{x_e,2} \cr
\psib_{x_e,2}
\end{array}
\right)
\  \mbox{and}  \ 
\left(
\begin{array}{c}
\psi_{x_o,1} \cr
\psib_{x_o,1} \cr
\psi_{x_o,2} \cr
\psib_{x_o,2}
\end{array}
\right)
\ \rightarrow \ 
 V^* \ \left(
\begin{array}{c}
\psi_{x_o,1} \cr
\psib_{x_o,1} \cr
\psi_{x_o,2} \cr
\psib_{x_o,2}
\end{array}
\right)
\end{equation}

where $x_e$ and $x_o$ refer to even and odd lattice sites resepectively and $V$ is a $SU(4)$ matrix in the fundamental representation.

The observables we wish to measure are the susceptibilities :
\label{susdefs}
\begin{eqnarray}
\chi_{1} &=& \frac{1}{2L^3}\sum_{x,y,x\neq y}\langle \overline{\psi}_{x,1}\psi_{x,1}\overline{\psi}_{y,1}\psi_{y,1} \rangle.
\\
\chi_{2} &=& \frac{1}{2L^3}\sum_{x,y,x\neq y} \langle \overline{\psi}_{x,1}\psi_{x,1}\overline{\psi}_{y,2}\psi_{y,2} \rangle.
\end{eqnarray}

where expectation values are defined as
\begin{equation}
\Big\langle {\cal O} \Big\rangle = \frac{1}{Z}
\int [d\overline{\psi}\ d\psi]\ {\cal O}\ \mathrm{e}^{-S(\overline{\psi},\psi)}
\end{equation}
with $Z$ being the partition function. The presence of a condensate can be inferred when these susceptibilities diverge as $L^3$ for large values of $L$. A constant susceptibility at large $ L $ would signal a zero condensate.  Using the above $ SU(4) $ symmetry, it can be shown that all other fermion bilinear susceptibilities can be expressed in terms of $ \chi_1 $ and $ \chi_2 $. Another observable that we measure is the local four-point condensate defined by
\begin{equation}
\rho_m = \frac{1}{L^3} \ \sum_x\ \langle \overline{\psi}_{x,1}\psi_{x,1}\overline{\psi}_{x,2}\psi_{x,2} \rangle.
\end{equation}

\section{The Fermion Bag Approach}
\label{sec4}

The traditional method for studying four-fermion field theories is by introducing an auxiliary field to convert the four-fermion term into a fermion bilinear. In this work we use an alternative Monte Carlo approach introduced a few years ago, called the fermion bag approach \cite{fb1}. A review of the fermion bag approach can be found in \cite{fb2}.

\begin{figure}
\centering
\parbox{7cm}{
\includegraphics[width=\linewidth]{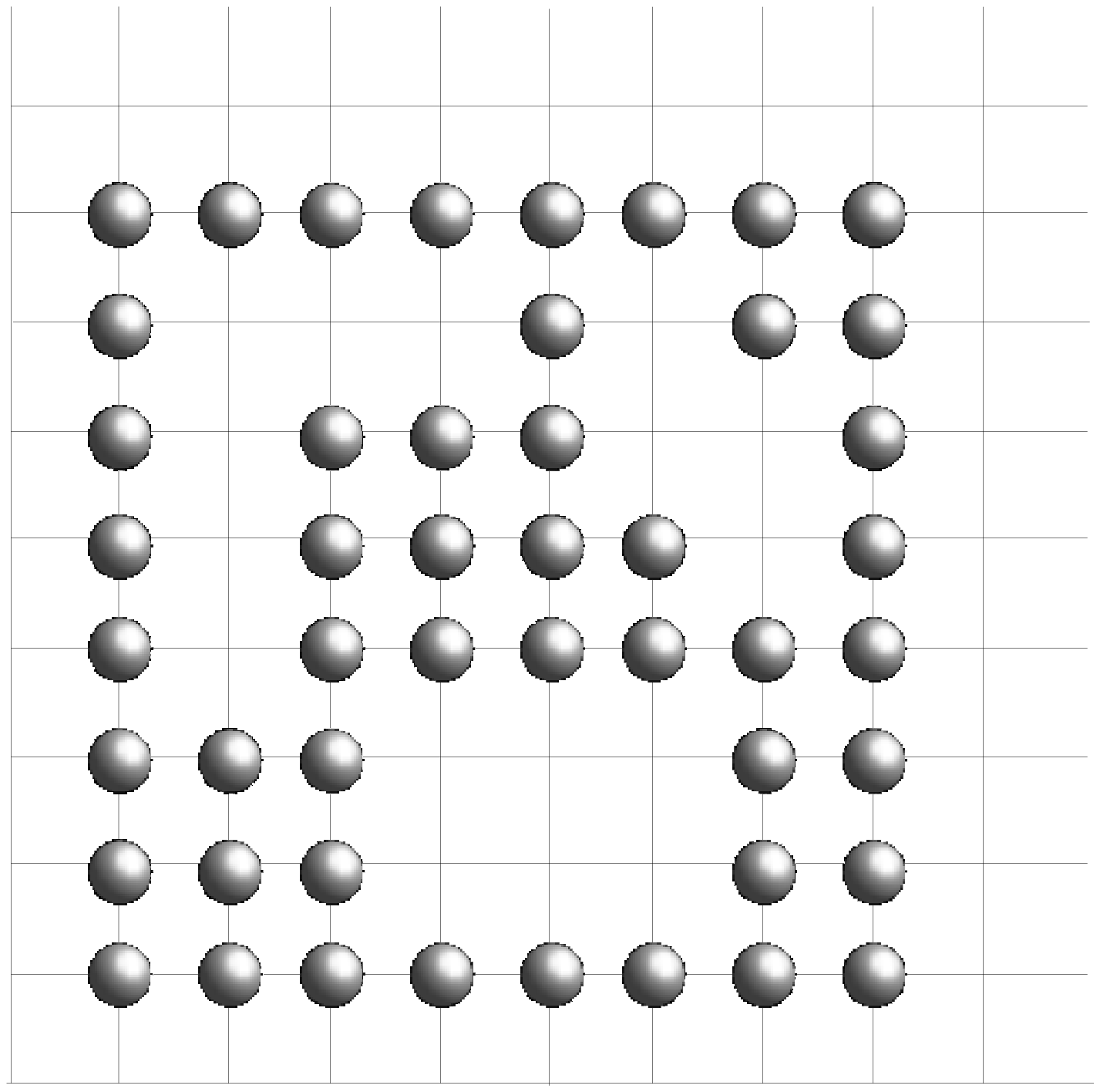}
\caption{\label{monoconf} An example of a monomer configuration $[n]$ showing free fermion bags on a two dimensional lattice. The blue circles denote the monomer sites.}}
\qquad
\begin{minipage}{7cm}
\includegraphics[width=\linewidth]{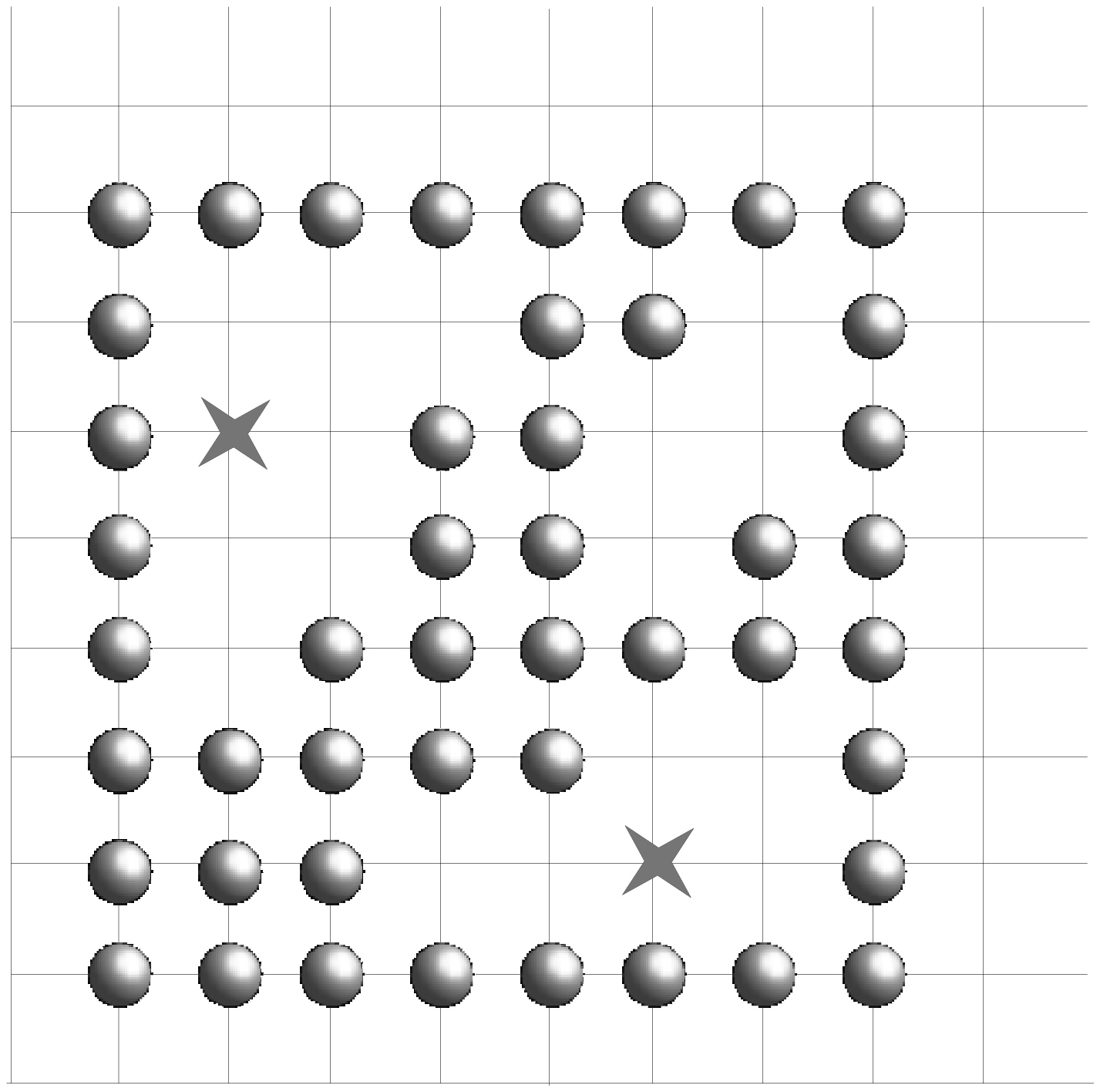}
\caption{ An example of a monomer configuration $[n]$  giving a zero contribution for the fermionic operator. The crosses denote the source points $ x $ and $ y $. }
\label{zerowt}
\end{minipage}
\end{figure}

\begin{figure}
\centering
\parbox{7cm}{
\includegraphics[width=\linewidth]{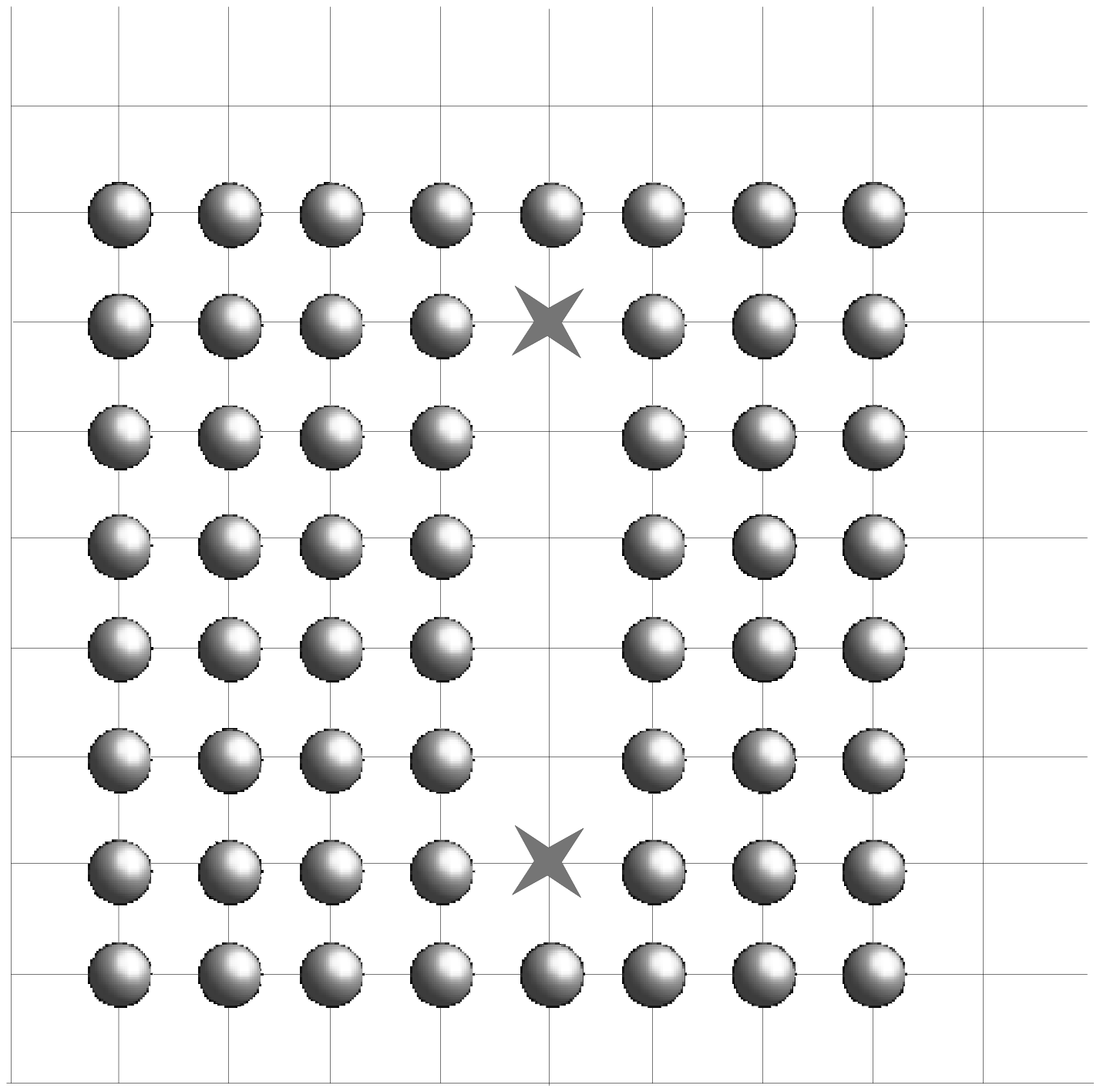}
\caption{\label{leadconf} The leading contribution with one fermion bag, at strong coupling. The crosses denote the source points $ x $ and $ y $. }}
\qquad
\begin{minipage}{7cm}
\includegraphics[width=\linewidth]{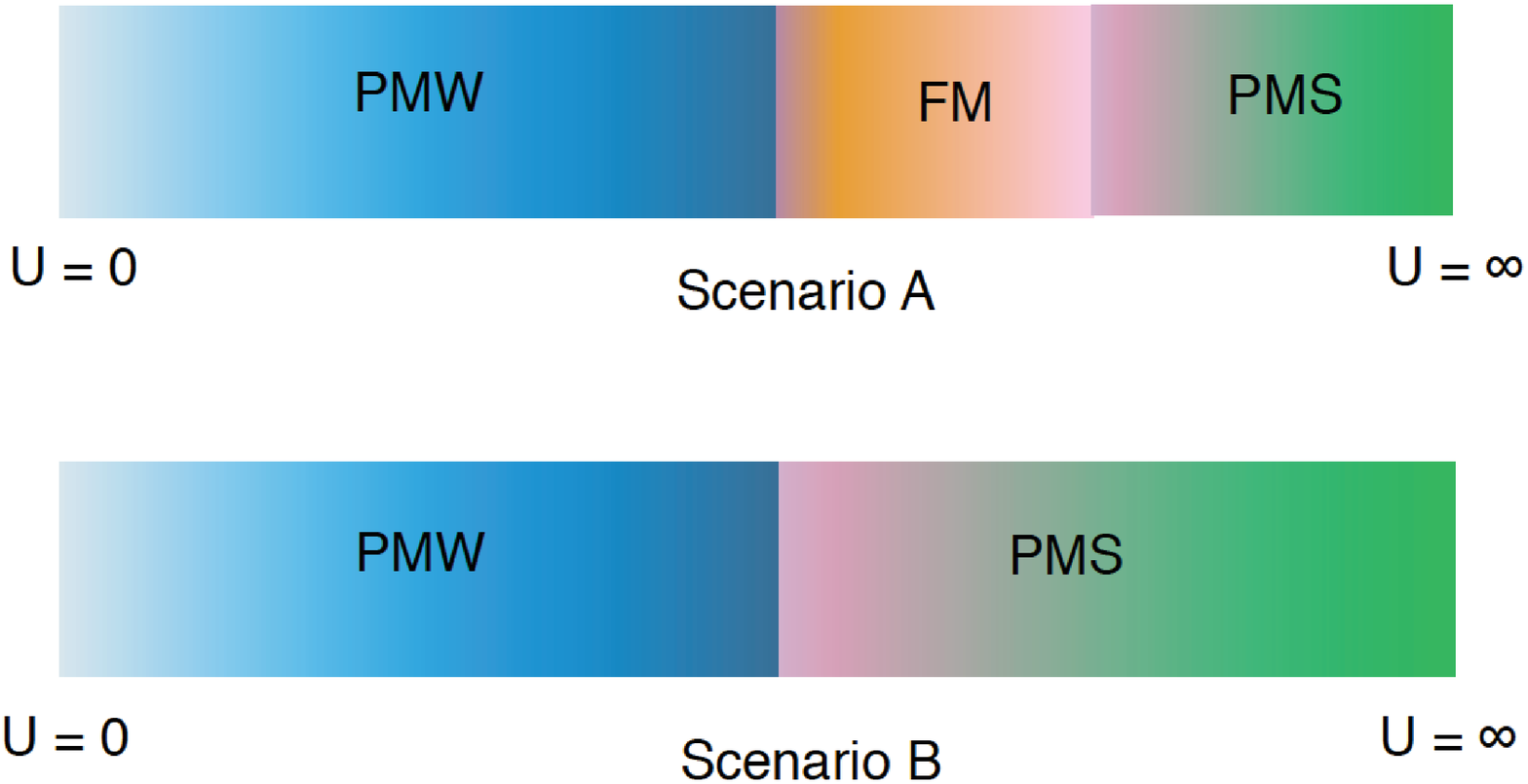}
\caption{\label{pdiag} The two possible phase diagrams for our model based on previous studies. Our work is consistent with scenario B with a second order transition between the PMW phase and the PMS phase.}
\end{minipage}
\end{figure}
In the Fermion bag approach, by defining a binary lattice field $n_x = 0,1$, we divide the lattice sites into monomers ($n_x=1$)  and free sites ($n_x=0$). Calling each such distribution a configuration $[n]$,  (Figure \ref{monoconf}) we can express our observables as a sum over these configurations. 
For example, the partition function can be written in the strong and weak coupling limits as follows:
\begin{eqnarray}
Z \ &=&\  \sum_{[n]}\ U^k \ \prod_{\cal B}\Big(\mathrm{Det}(W_{\cal B})\Big)^{2}
\\
Z \ &=&\  \Big(\mathrm{Det}(M)\Big) \sum_{[n]}\ U^k \ \Big(\mathrm{Det}(G)\Big)^{2},
\end{eqnarray}
where $k$ represents the number of monomers in the configuration, $M$ is the free staggered fermion matrix defined in (\ref{staggered}), $W_{\cal B}$ represents a free staggered fermion matrix connecting the sites within the bag ${\cal B}$, while $G$ represents a $k \times k$ free staggered propagator matrix connecting monomer sites.

There are two equivalent ways to look at fermion bags : 
\begin{itemize}
\item[1] In the weak coupling limit, we can think of fermions as living on the monomer sites and being able to hop on to other monomer sites through the free fermion propogators. The collection of all monomer sites can be thought of as a fermion bag.
\item[2] In the strong coupling limit, we can think of fermions as living on the free sites and being able to hop on to other free sites due to the free action. Each group of connected free sites can be thought of as forming a fermion bag. 
\end{itemize}

\subsection{ Phases }
\label{phase}

Using the Fermion bag approach, it is easy to visualize the different phases of the model in the strong and weak coupling limits. 
In 3D, since the four-fermion coupling is irrelevant, we must have a massless phase at weak couplings. 
To understand the behavior at strong couplings, let us write the fermionic correlator in the form:
\begin{eqnarray}
\langle \psi^1_x \overline{\psi^1_y} \rangle &=& \frac{1}{Z} \sum_{[n]} \ U^{N_m} \ \prod_{\cal{B}} \left( \det(W_B) \right)^2  {W^{-1}}_{{\cal{B}};x,y} \label{fcorr}
\end{eqnarray}
 where  $ {W^{-1}}_{{\cal{B}};x,y} $ is the inverse of the Dirac operator within the free fermion bag $ \cal{B} $ containing the points $  x $ and $ y $. 
For large $ U $, the leading contribution will come from configurations with very few free sites and hence the free fermion bags will be small. Since the free action only connects nearest-neighbors, fermions cannot hop between different bags. Hence if the sites $ x $ and $ y $ where the fermion sources are inserted belong to different bags, (  Fig \ref{zerowt}) then  (\ref{fcorr}) requires the fermionic correlator to be zero.
Hence, the leading contributions will come from configurations having just one fermion bag containing both points $ x $ and $ y $ ( Fig \ref{leadconf} ). Since large fermion bags are exponentially supressed, the correlator decays exponentially, implying a massive phase at large couplings. Similarly, using a slightly different argument we can show that the bosonic correlator also decays exponentially at large couplings \cite{our}. This implies that the condensate is zero.

As mentioned in the Introduction, such exotic mechanisms of fermion mass generation have appeared in literature in the context of Yukawa models. Mean field calculations have shown a phase transition from a massless phase ( referred to as PMW or weak paramagnetic phase ) to an intermediate massive phase with a fermion bilinear chiral condensate ( referred to as FM phase ) and another transition from this intermediate phase to a new massive phase with zero fermion bilinear condensates ( referred to as PMS phase )\cite{p2}. 
Interestingly, other mean field calculations \cite{p3,p4} for 2 flavors in 3D give a single first-order phase transtion from the PMW to the PMS phases. These two scenarios are shown in Fig \ref{pdiag}. In our work, we find a single second-order transition from the  PMW to PMS phases.

\section{ Analysis and results}
We have collected data for cubical lattices with sizes ranging from $ 12^3 $ upto $ 28^3 $, using three different Monte-Carlo algorithms. A detailed description of our algorithms and analysis can be found in \cite{our}.

\begin{figure}[!htb]
\centering
\parbox{7cm}{
\includegraphics[width=\linewidth]{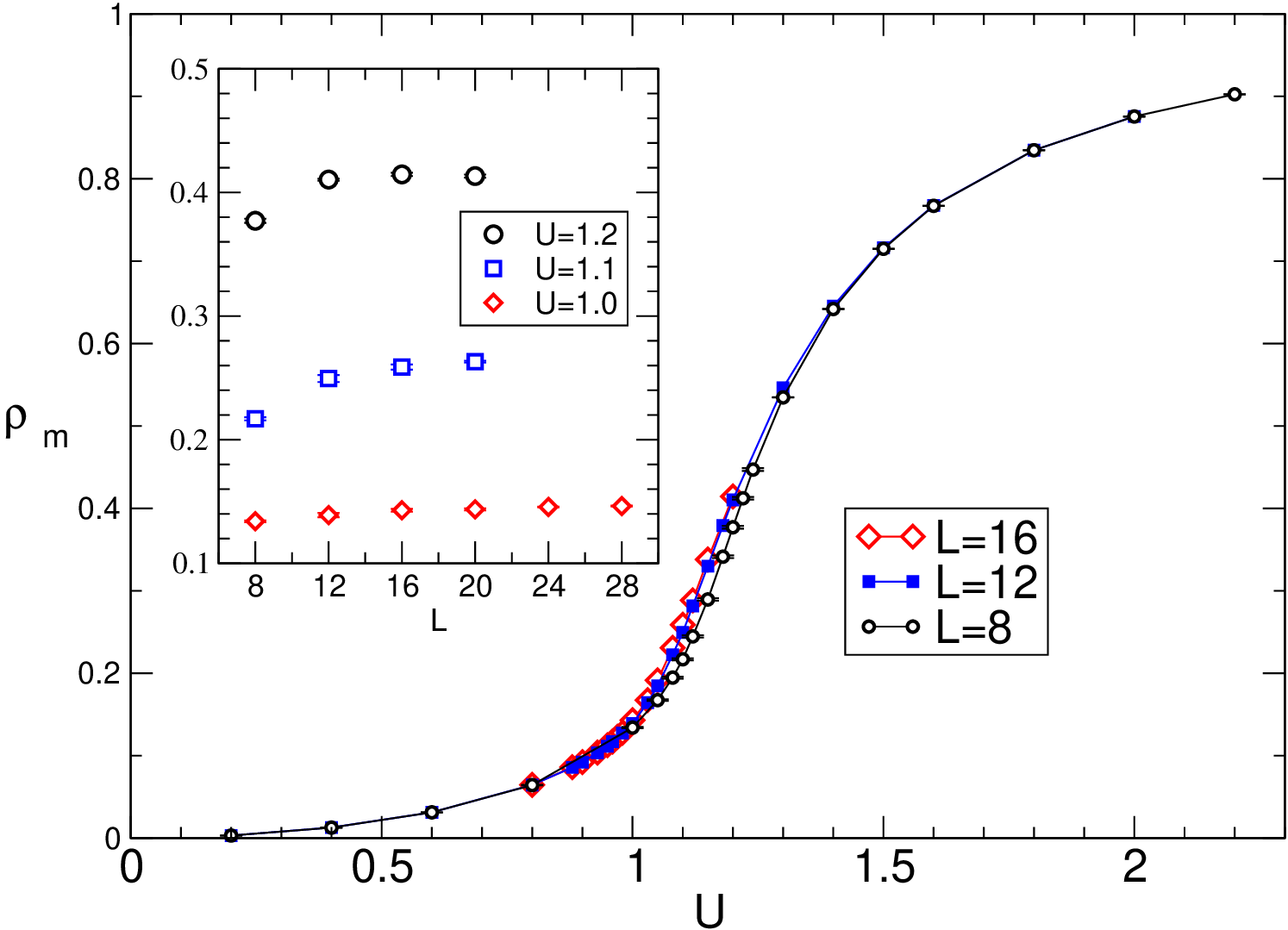}
\caption{\label{rhomono} Variation of the average monomer density $ \rho_m $ as a function of coupling $ U $ for lattices of size 8,12,16.}}
\qquad
\begin{minipage}{7cm}
\includegraphics[width=\linewidth]{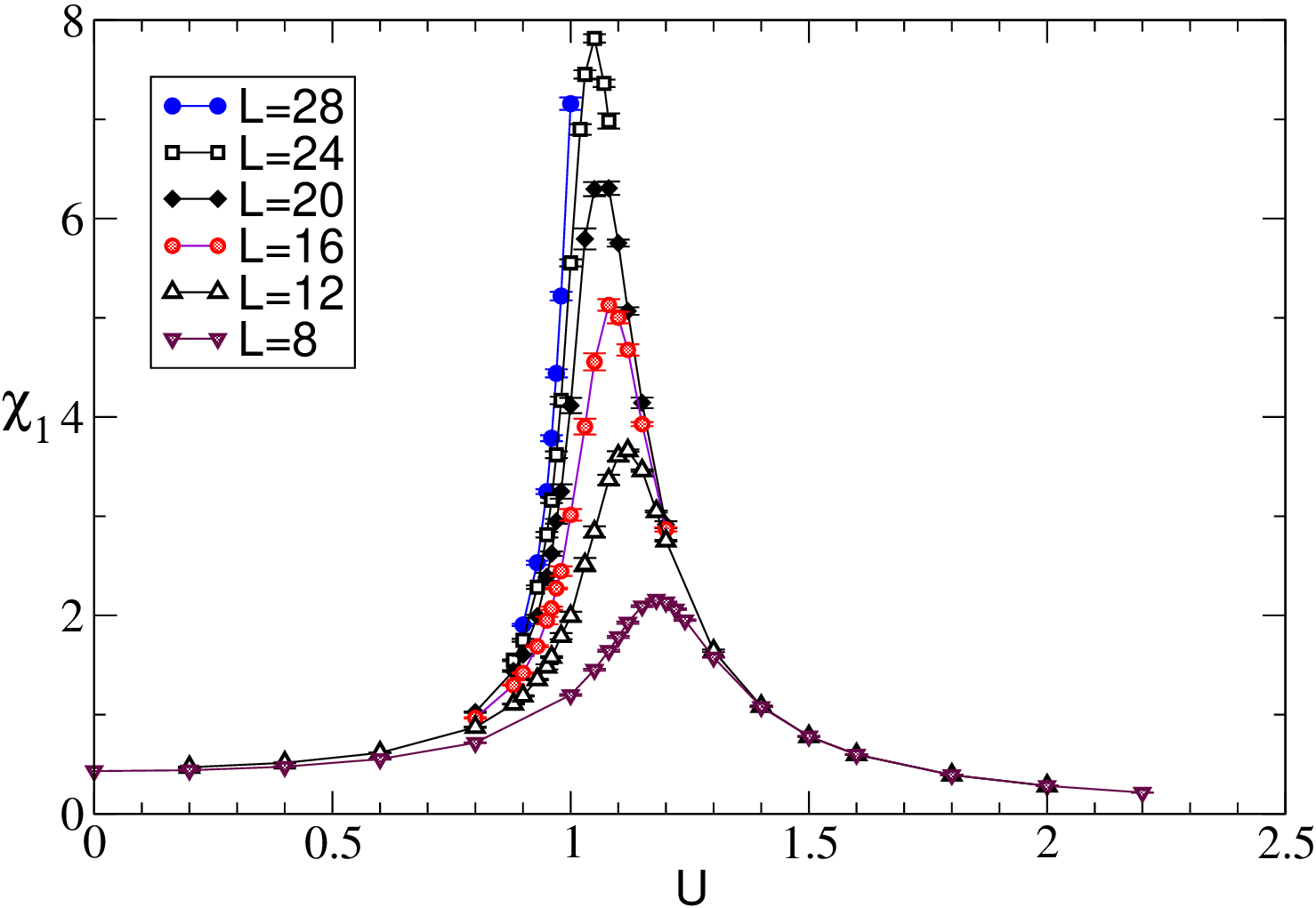}
\caption{\label{chi_v_U} Variation of the susceptibility $ \chi_1 $ as a function of coupling $ U $.}
\end{minipage}
\end{figure}

\begin{figure}[!htb]
\centering
\parbox{6.5cm}{
\includegraphics[width=\linewidth]{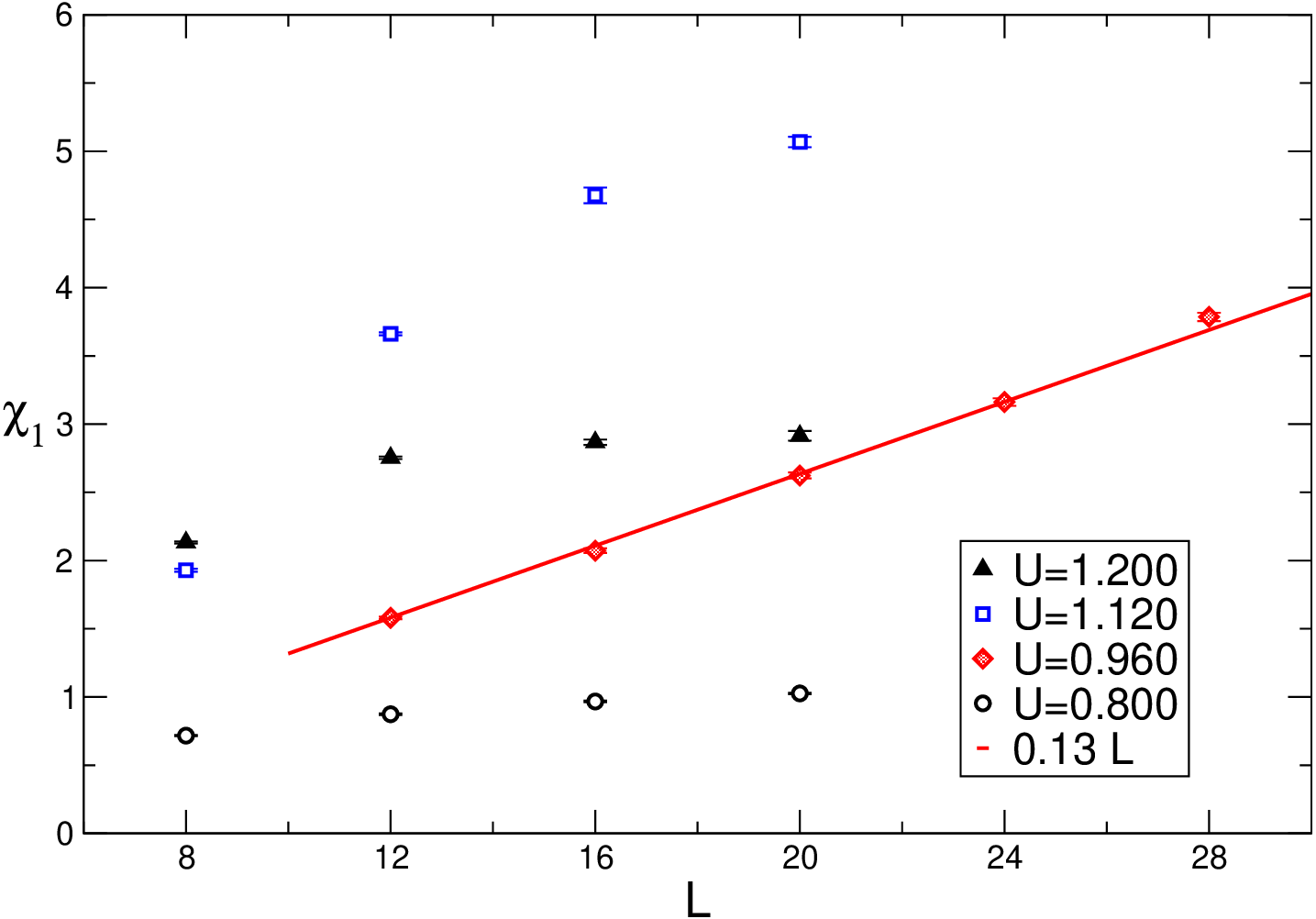}
\caption{\label{chi_v_L} Variation of the susceptibility $ \chi_1 $ as a function of lattice size $ L $. Note the linear rise of the susceptibility for $ U = 0.96 $.}}
\qquad
\begin{minipage}{7.5cm}
\includegraphics[width=\linewidth]{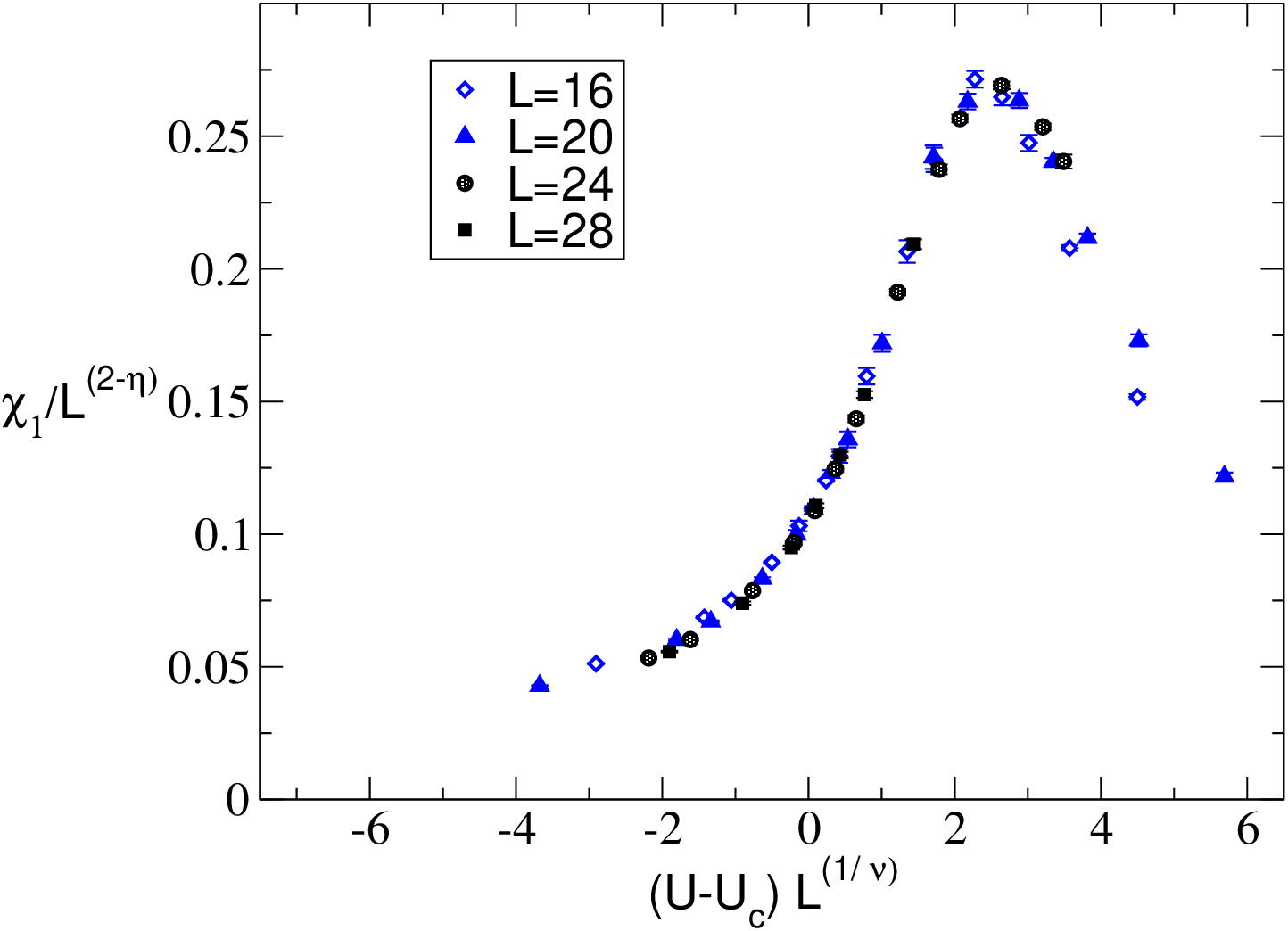}
\caption{\label{trans_order} Evidence for a second order phase transition}
\end{minipage}
\end{figure}

Looking at the behavior of the four-point condensate $ \rho_m $ as a function of the coupling $ U $ in Figure \ref{rhomono} it is clear that it is a smooth function, increasing from 0 for small couplings, rising sharply near $ U \sim 1 $ and approaching 1 for large couplings. The thermodynamic limit is reached by L=16.
Figure \ref{chi_v_U} shows the behavior of the susceptibility $ \chi_1 $ as a function of the coupling $ U $ for various values of the lattice size $ L$ . The susceptibility is a smooth function of $ U $, reaching a maximum for $ U \sim 1 $.  If we look  at the behavior of $ \chi_1 $ as a function of $ L $ for different values of  $ U $ in Figure \ref{chi_v_L},  it can be seen that the susceptibility saturates with lattice size for both small and large couplings. The lack of an $ L^3$ divergence implies the absence of a condensate. However near $ U = 0.96 $, the susceptibility increases linearly with $L$, which is consistent with a second order critical point.
The susceptibility $ \chi_2 $ also shows a similar behavior.
All this points to a single phase-transition in the region close to $ U_c = 0.96 $.

We have used scaling relations of both susceptibilities and performed a combined fit to extract the critical exponents. We present two such fits in table \ref{fits}.
\begin{table*}[!htb]
 \centering
\begin{tabular}{|c||c|c|c|c|}
\hline
Fit & $ U_c $ & $ \nu $ & $ \eta $ & $ \chi^2  $ \\
\hline
\hline
1 & 0.957(1) &  0.95(5) & 0.940(9) & 1.1 \\ 
2 & 0.958(1)  & 1.24(2)  & 0.884(1) & 1.9\\
\hline
\end{tabular}
\caption{ Critical exponents obtained using two different fits }
\label{fits}
\end{table*}

Thus, although the critical point seems to be constrained well enough, the exponents $ \nu $ and $ \eta $ are not. 
This suggests the need for data on larger lattices to constrain the exponents.

From the theory of second-order phase transitions, near a second-order critical point, we know that the susceptibility should vary continuously i.e.
\begin{equation}
  \chi \sim {L^{2-\eta}} f\left( (U-U_c)L^{\frac{1}{\nu}} \right) 
\end{equation}
Hence, a plot of $ \frac{\chi}{L^{2-\eta}} $ vs $ (U-U_c)L^{\frac{1}{\nu}} $ should be a smooth function.
Figure \ref{trans_order} shows that this is indeed true. This points to a second-order phase transition between the PMW phase and the PMS phase.

\section{ Conclusions and Future work}
We have studied a lattice model in three Euclidean dimensions in which fermions can acquire a mass, but without a fermion bilinear condensate.
The transition from massless to massive phase seems second order. This suggests the possibility of defining a continuum quantum field theory at the critical point.
To obtain the precise values of the critical exponents, we plan to extend this calculation to larger lattices. 
We also plan to extend this work to 4D to explore if a similar phase transition exists there.

\end{document}